\documentclass[%
 reprint,
 amsmath,amssymb,
 aps,
]{revtex4-2}

\usepackage{graphicx}
\usepackage{dcolumn}
\usepackage{bm}
\usepackage{amsmath}
\usepackage{amssymb}
\usepackage{hyperref}
\usepackage[mathlines]{lineno}
\usepackage{float}

\begin{document}

\preprint{APS/123-QED}

\title{Least-Dissipation Interfaces in Fully Miscible Fluids}

\author{Gyeong Min Choi}
 \affiliation{Department of Chemical Engineering (BK-21 Four), Dong-A University, Hadan 840, Saha-gu, Busan 49315, Republic of Korea}
\author{Heon Sang Lee}%
 \email{Contact author: heonlee@dau.ac.kr}
\affiliation{Department of Chemical Engineering (BK-21 Four), Dong-A University, Hadan 840, Saha-gu, Busan 49315, Republic of Korea}

\date{\today}

\begin{abstract}
Sharp interfaces in miscible fluids have long been observed, yet classical theory associates them only with phase coexistence and non-convex free energies. We present a minimal variational framework where adding a Fermi–Dirac (FD) free energy to a convex bulk contribution preserves miscibility, while the convex term drops out in the gradient balance, leaving the FD part to set a logistic profile by the principle of least dissipation. After formation, the FD interface propagates diffusively and later broadens by ripening, giving a unified view of interfacial dynamics in miscible fluids.
\end{abstract}

\maketitle

Sharp interfaces in fully miscible fluids have long been observed — from the pioneering works of Korteweg ~\cite{ref1} and Joseph ~\cite{ref2} to modern microchannel experiments ~\cite{ref3,ref4,ref5,ref6,ref7,ref8,ref9,ref10,ref11,ref13,ref14,ref15} demonstrating a critical Péclet threshold near 850 ~\cite{ref16,ref17,ref18,ref19,ref20}. Yet despite being experimentally well established, these interfaces have lacked a rigorous theoretical foundation under convex thermodynamics. Recent studies have examined how such interfaces persist under diffusion and whether nonequilibrium interfacial tension can impart effective elasticity ~\cite{ref16,ref17,ref21,ref22}. In core–sheath microchannel flows, an apprently immiscible interface can remain stable above the critical Péclet number, while halting the core stream under continuous sheath flow leads to a wet capillary thinning (WCT) regime that directly exposes interfacial mechanics ~\cite{ref16,ref17}. Beyond microfluidics, examples from marine, biological, and polymeric flows further illustrate that sharp interfaces can persist even without bulk phase separation, playing important roles in micro rheology ~\cite{ref23,ref24,ref25,ref26,ref27}, microvascular blood flow ~\cite{ref28,ref29}, polymer blend morphology ~\cite{ref30}, colloidal stability ~\cite{ref31,ref32}, targeted drug delivery ~\cite{ref33,ref34}, and pollutant dispersion ~\cite{ref35}.

Despite extensive research, the mechanism by which sharp interfaces persist in fully miscible fluids without phase separation remains unresolved. Classical phase-field models based on double-well free-energy landscapes require bulk phase instability and coexisting phases—conditions absent in these steady-state experiments ~\cite{ref36,ref37,ref38,ref39,ref40,ref41,ref42,ref43,ref44}. While such models capture equilibrium phase separation, they cannot explain the localized, dynamically sustained interfaces observed in miscible systems. This gap persists because the dominant paradigm for sharp interfaces has been rooted in non-convex free-energy formulations tied to phase coexistence, leaving convex free-energy systems—typical of fully miscible fluids—largely unexplored. Moreover, the consistency between local chemical-potential equilibrium and sustained mass flux, and its implications for interfacial dissipation, has not been addressed within a rigorous analytical framework.

Here we show that, within a square-gradient formulation, one can derive a free-energy functional whose Euler–Lagrange equation yields a Fermi–Dirac logistic profile exactly. This profile embodies Onsager’s least-dissipation path, minimizing entropy production at the instant miscible fluids come into contact. Driven by bulk chemical potential, the interface subsequently ripens as a Lyapunov-stable route toward global equilibrium. In contrast to classical models requiring double-well free energy, our formulation shows that even a convex free energy necessarily produces a sharp interface [Fig. 1(a)]. Because convexity is usually taken as a guarantee of smooth mixing without interfaces, this overturns a long-standing assumption. Thus, the spontaneous emergence of interfaces in miscible fluids appears not as an anomaly, but as the natural outcome of the principle of minimum entropy production. Although the total bulk free energy is convex due to the activity term, at the moment of interface formation the activity contribution, being a purely local function of \textit{c}, drops out with the balance under the principle least dissipation, leaving the FD quartic to determine the interfacial structure.

To formalize this least-dissipation principle, we introduce the global potential density, $\omega(c) = f(c) - \mu_{0} c$, which yields a logistic profile exactly when $\omega(c)$ is quartic in $c$, without requiring the double-well form of classical Landau theories. We then construct the steady transport using the Rayleighian~\cite{ref45}

\begin{equation}
\mathcal{R}[J,\mu] 
= \left( \frac{1}{T} \right) \int J \, (-\partial_{z}\mu) \, dz 
- \left( \frac{1}{2T} \right) \int \frac{J^{2}}{M(c)} \, dz,
\end{equation}
whose minimization yields Onsager’s law $J = -M(c)\,\partial_{z}\mu \;\text{ with }\; \partial_{z}J = 0$. Combining with $\mu = f'(c) - \kappa \partial_{z}^{2} c.$ gives

\begin{equation}
\kappa c'' = f'(c) - \mu_{0}, 
\tag{2}
\end{equation}
which is solved exactly by the logistic profile [Fig. 1(b)]

\begin{equation}
c_{F.D.}(z) = \left[ 1 + \exp\!\left( \frac{z - z_{0}}{\delta} \right) \right]^{-1},
\tag{3}
\end{equation}
corresponding to a quartic free energy
\begin{equation}
f_{F.D.} = \mu_{0}c + \Lambda \left( \tfrac{1}{2}c^{2} - c^{3} + \tfrac{1}{2}c^{4} \right), 
\quad \Lambda = \frac{\kappa}{\delta^{2}} .
\tag{4}
\end{equation}

In this least-dissipation state, the chemical-potential gradient vanishes inside the interface ($\partial_{z}\mu \equiv 0$), so the entire thermodynamic drop $\Delta \mu$ is expelled into the bulk, and the interfacial entropy production tends to zero, $\Sigma_{FD} \to 0$  as $t \to 0$. Using the same $c(z)$, $\mu(z) = \mu_{0}$ for FD-only, while including $g(c)$ makes $\mu(z)$ linear across the interface [Fig. 1(d)]. Thus, at the very moment of contact between miscible fluids, the characteristic Fermi–Dirac–type profile  necessarily emerges.

\begin{figure}[htbp]
\centering
\includegraphics[width=1\linewidth]{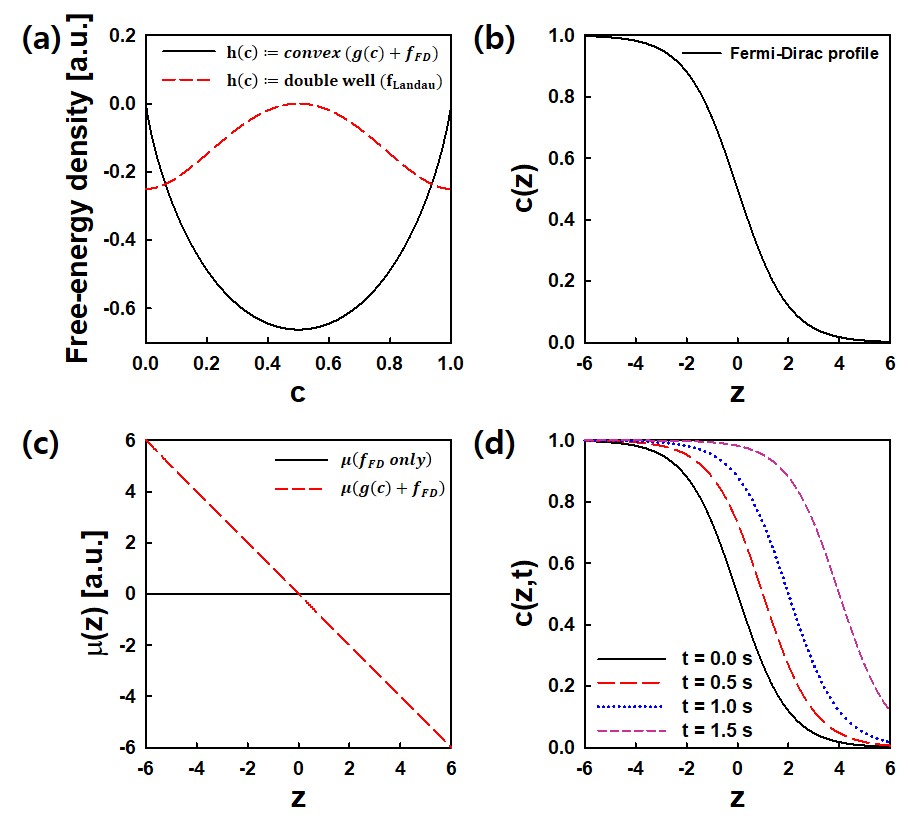}
\caption{\label{fig:epsart}(a) Comparison of free-enery densities. The present convex model is $h(c) = g(c) + f_{FD}$. For reference, the classical Landau free energy is $f_{landau}(c) = -A_{L}/2 \, \varphi^2 + B_{L}/2 \, \varphi^4$, 
    $\varphi = 2c-1$. The convex $h(c)$ remains bounded and strictly positive in curvature, while $f_{landau}$ exhibits the canonical doulle-well structure. (b) Fermi-Dirac (FD) interface with $\delta = 1 \, \mu m$. (c) Using the same $c(z)$, the FD-only chemical potential is flat $\mu_z = \mu_0$, whereas including the ideal bulk free energy $g(c) = RT[c \ln c + (1-c)\ln(1-c)]$ yields $\mu(z) = \mu_0 + RT[c/(1-c)] = \mu_0 - (RT/\delta)(z-z_0)$, i.e., linear across the interface. (d) Propagating FD interface $c(z,t) = c(z-Vt)$.}
\end{figure}

Having established the variational origin of the Fermi–Dirac interface, we next examine its dynamical stability by embedding the profile in the Cahn–Hilliard dynamics ~\cite{ref36} and analyzing the associated Lyapunov functional, which reveals the conditions under which the interface thickens ($\delta$) and ultimately ripens toward global equilibrium. Let us now consider the free-energy functional $F[c]$, which plays the role of a Lyapunov functional for the Cahn–Hilliard dynamics ~\cite{ref36} and governs the ripening of the interface

\begin{equation}
F[c] = \int_{-\infty}^{\infty} \left[ g(c) + f_{FD}(c) 
+ \frac{\kappa}{2} \left( \partial_z c \right)^{2} \right] dz,
\tag{5}
\end{equation}
here $g(c)$ denotes the bulk free energy, such as the ideal convex form $g(c) = RT[c \ln c + (1-c)\ln(1-c)]$. though it need not necessarily be ideal. At the moment of interface formation, the chemical potential must be spatially uniform ($\partial_{z}\mu=0$). Since the activity term from $g(c)$ depends only on c and is already homogeneous across the interface, its contribution cancels in the gradient balance. As a result, the interfacial profile is governed solely by the FD component, effectively $h=f_{FD}$. Since $g(c) \gg f_{FD}(c)$ in the limit $(\partial_z c)^2 \to 0$, the bulk free energy acts as the global driving force toward equilibrium, while $f_{FD} (c)$ together with the gradient term sustains the interfacial structure. $F(c)$ acts as a Lyapunov functional of the Cahn–Hilliard dynamics ~\cite{ref36}, since ${dF}/{dt} \leq 0$. With $\mu \equiv h'(c) - \kappa \partial_z^2 c$ and $h(c) = g(c) + f_{FD}(c)$, the second variation $\delta^2 F(\eta) \geq 0$ ensures Lyapunov stability, the only neutral mode of the associated operator $\mathcal{L} = -\kappa \partial_z^2 + h''(c)$ corresponds to translation of the interface, consistent with the convexity condition  $h''(c) > 0$. If $g(c)$ were absent, the dynamics would not be Lyapunov stable, since $f_{FD}(c)$ is not convex over the entire composition range. Thus, the ripening of the interface is ultimately driven by the convex bulk free energy $g(c)$.

We next compare the interfacial thickening dynamics with the classical Fickian case. According to Onsager’s principle of least dissipation ~\cite{ref46,ref47}, the evolution of the interfacial width follows

\begin{equation}
\dot{\delta} \sim -\frac{1}{R(\delta)} \frac{\partial F}{\partial \delta} .
\tag{6}
\end{equation}

In the absence of the bulk free energy $g(c)$, the relaxation of the Fermi–Dirac–type interface is spontaneous, $\delta(t)\sim~t^{1/4}$ in the interface-limited regime and $\delta(t)\sim t^{1/3}$ in the bulk-limited regime. When driven by the global force of $g(c)$, the thickening laws are modified to  

\begin{equation}
\delta(t) \sim \delta_{0} \exp\!\left[ \frac{(\Delta \mu)^{2}}{\alpha \kappa} \, t \right] 
\quad \text{(interface-limited)},
\tag{7}
\end{equation}

\begin{equation}
\delta(t) \sim \left[ \delta_{0}^{-1} - \frac{(\Delta \mu)^{2}}{\kappa R_{\textit{bulk}}} \, t \right]^{-1} 
\quad \text{(bulk-limited)}.
\tag{8}
\end{equation}

Eqs. (7)–(8) serve as scaling/local forms [Eq. (8) assumes a local constant $R_{bulk}$. The exact driven ripening law with time-dependent $R_{bulk}(t)$ is given in the Supplemental Material ~\cite{ref48}, Eq. S$29^{\prime}$ (with $t_{s}\equiv t_{1}$), and is used for quantitative evaluations and for $t_{2}$ via slope matching. These expressions show that the presence of the bulk free energy $g(c)$ accelerates the ripening, in stark contrast to the much slower spontaneous relaxation of the undriven Fermi–Dirac interface. In contrast, pure Fickian diffusion yields $\delta(t)\sim t^{1/2}$, accompanied by a divergent entropy generation $\Sigma_{\text{FICK}} \to \infty$ as $t \to 0$, highlighting the fundamentally different dissipation pathway from the Fermi–Dirac interface. By comparison, the FD interface without $g(c)$ exhibits vanishing interfacial entropy generation $\Sigma_{FD} \to 0$ as $t \to 0$, while inclusion of the bulk free energy $g(c)$ in the square-gradient functional leads to a finite total entropy generation $\Sigma_{\textit{tot}}$, which drives the accelerated ripening toward equilibrium.

\begin{equation}
\Sigma_{\textit{tot}}(0^{+}) 
= \frac{RT}{R_{\textit{tot}}(\delta_{0})}\frac{(\Delta c)^{2}} {[c_{0}(1-c_{0})^{2}]}.
\tag{9}
\end{equation}

For Fig. 2, we show the simultaneous evolution of the interfacial thickness $\delta(t)$ and the interfacial entropy production $\Sigma_{\textit{int}}(t)$. Transition times $t_{1}$ (FD$\to$ripening) and $t_{2}$ (ripening$\to$Fickian) are obtained analytically by equating interfacial and bulk resistances and by inner–outer matching; closed forms are given in the Supplementary Material ~\cite{ref48}, Eqs. (S85) and (S87) [shape factors in Eq. (S83), with $\chi = \pi$ and $\lambda = a\sqrt{\pi}/2$]. The transition conditions $t_{1}$ marks the onset of dissipation, where $\delta(t)$ departs from the FD $t^{1/4}$ scaling and $\Sigma_{\textit{int}}$ jumps from zero to a finite value once the interfacial and bulk resistances balance, while $t_{2}$ marks the crossover to classical Fickian diffusion when the FD thickness matches the outer diffusive layer and the intermediate ripening regime is characterized by $\delta(t)\sim(t-t_{1})^{1⁄3}$ and $\Sigma_{int}(t)\sim(t-t_{1})^{-1/3}$, followed by the Fickian regime with $\delta(t)\sim t^{1⁄2}$ and $\Sigma_{int}(t)\sim t^{-1/2}$. The parameter a represents an effective thickness calibration factor that can be determined experimentally from slope matching in core–sheath profiles. From a purely theoretical standpoint one may set a as an order-unity constant, but its precise value is system dependent and best calibrated against experiments. The dependence on $a$ differs for the two times, since $t_{1}$ is independent of $a$ whereas $t_{2}$ varies explicitly with a through $\lambda$, the slope-matching coefficient that connects the inner logistic profile to the outer error-function tail.

Thus, bulk free energy not only stabilizes but also accelerates the interfacial broadening toward equilibrium. We now turn from broadening to translation, considering how the FD interface propagates under a finite driving force.

\begin{figure}[hbpt]
\centering
\includegraphics[width=0.8\linewidth]{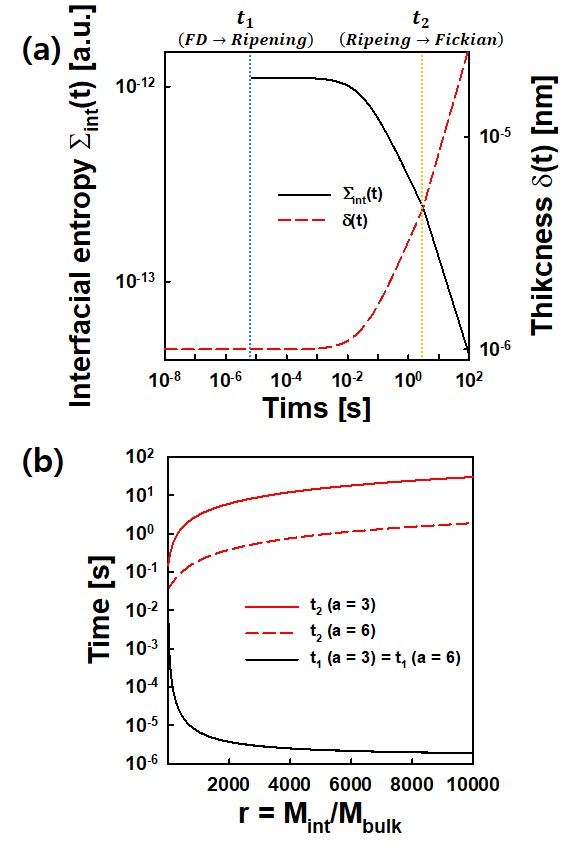}
\caption{\label{fig:epsart}Simultaneous evolution of interfacial thickness $\delta(t)$ and entropy generation $\Sigma_{int}(t)$. In the FD regime, $\delta\sim t^{1/4}$ while $\Sigma_{int}=0$. At $t_{1}$,entropy generation jumps up and the ripening regime follows with $\delta\sim(t-t_{1})^{1/3}$ and $\Sigma_{int}\sim(t-t_{1})^{-1/3}$. At $t_{2}$, the crossover to Fickian diffusion occurs, where $\delta\sim t^{1/2}$ and $\Sigma_{int}\sim t^{-1/2}$ (Representative parameters: $\delta_{0}=1\,\mu\text{m}$, $\kappa = 10^{-12}$, $\chi = \pi$, $a = 3$, $\lambda = a\sqrt{\pi}$, $M_{\textit{bulk}} = 10^{-12}\,\text{m}^{2}/\text{s}$, $\Gamma = 1$). (b) Transition times vs mobility ratio $r = M_{\textit{int}}/M_{\textit{bulk}}$. Closed-form $t_{1}$ (FD$\to$ripening) and $t_{2}$ (ripening$\to$Fickian) from the Supplemental Material ~\cite{ref48}, Eqs. (S85), (S87) with shape factors from (S83) with the effective thickness calibration factor $a=3$ and $a=6$. Ripening in (a) is visualized with a local constant-resistance form consistent with Eq. (8); the exact time-dependent bulk-resistance law is given in SM Eq. (S29$^{\prime}$) with $t_{s}\equiv t_{1}$. Transition times in (b) follow the Supplemental Material ~\cite{ref48}, Eqs. (S85) and (S87); for non-ideal $g(c)$, use $D_{\textit{eff}} \to D_{\textit{eff},g}$ and $\lambda \to \lambda_{g}$.}
\end{figure}

Under a small imposed chemical-potential difference $\Delta\mu$, the Cahn–Hilliard dynamics ~\cite{ref36} with the FD free energy admits a propagating-interface solution,

\begin{equation}
c(z,t) = c_{0}(z - Vt) + O(\varepsilon), \qquad V = {J}/{\Delta c}.
\tag{10}
\end{equation}

Linearization reveals a single translational zero mode and a finite spectral gap, so that the FD profile persists as a neutrally stable non-equilibrium steady state. In this regime, the interfacial chemical-potential gradient remains $O(\varepsilon^{2})$, dissipation vanishes to leading order, and the full thermodynamic drop $\Delta \mu$ is expelled into the bulk. The net flux is therefore governed solely by the bulk effective diffusivity, with the FD interface acting as a transparent kinetic barrier under steady propagation. This exact least-dissipation structure has no counterpart in double-well free energies, marking a distinct regime of miscible-fluid interfaces.

Strict Lyapunov stability, however, requires inclusion of the convex bulk free energy $g(c)$. With $h(c)=g(c)+f_{FD}(c)$, the dynamics acquires global convexity, ensuring entropy increase and eventual ripening. At long times and finite driving, the bulk free energy imparts a small but nonzero interfacial resistance, leading the FD interface to broaden while still propagating. This establishes propagation and ripening as complementary nonequilibrium pathways of miscible interfaces.

The same minimal-dissipation property appears in the effective-diffusivity representation,

\begin{equation}
J = -D_{\textit{eff}} \, \partial_{z} c,\ D_{\textit{eff}} = D \bigl[ 1 + c \, \partial_{c} \ln \gamma \bigr],
\tag{11}
\end{equation}
where $\gamma$ is the activity coefficient. For osmotic mixtures this reduces to ~\cite{ref16,ref17}

\begin{equation}
D_{\textit{eff}} \approx \frac{D v_{m} \Pi}{RT}.
\tag{12}
\end{equation}

with $\Pi$ the osmotic pressure, showing that osmotic compressibility rather than thermal diffusion controls the transport rate. Thus, the FD interface constitutes the exact least-dissipation structure of fully miscible fluids, with transport governed entirely by osmotic compressibility as evidenced by experiments ~\cite{ref16,ref17}.

The formation of sharp interfaces in miscible fluids with sustained propagating behavior has been widely observed in microchannel core–sheath flows. When the total flow rate is low, the sheath fluid—typically a solvent with higher chemical potential—diffuses into the core, leading to inflation of the core thread ~\cite{ref5,ref7,ref9,ref16,ref17,ref18,ref19}. Although the fluids are fully miscible, a distinct interface remains visible at intermediate flow rates. At very low flow rates, this interface gradually disappears as full mixing occurs. As the flow rate increases, the core diameter shrinks, and above a critical velocity, a sharp interface is stabilized and convected downstream ~\cite{ref5,ref7,ref9,ref16,ref17,ref18,ref19}. Notably, the critical Peclet number is consistently observed near $(Pe)_{c}\sim 850$ across a wide range of systems, including molecular–molecular and polymer–solvent or blood–solvent combinations ~\cite{ref16,ref17,ref18,ref19}. In the case of complex fluids, this threshold is accurately predicted by 

\begin{equation}
(Pe)_{c} = \frac{U_{c}h}{D_{\textit{eff}}},
\tag{13}
\end{equation}
where the effective diffusivity includes an osmotic drift term. This provides strong experimental support for the theoretical model. Below, we show that this critical Peclet number can be captured by a simple scaling argument.

A scaling estimate for the critical Peclet number can be obtained by comparing the diffusive and convective time scales across the interfacial thickness $\delta$. The diffusive time over $\delta$ is $t_{D}\sim\delta^{2}/D$, while the convective time across the domain of size $h$ is $t_{U}\sim h⁄U$. Requiring that diffusion across the interface be slower than convection yields

\begin{equation}
(Pe)_{c} \sim \left( \frac{h}{\delta} \right)^{2}.
\tag{14}
\end{equation}

Using typical microfluidic parameters ($h\sim30 \ \mu m$, $\delta\sim1 \ \mu m$), we obtain $(Pe)_{c}\sim 900$, in excellent agreement with experimental observations near $Pe \approx 850$. This quantitative agreement between theory and experiment confirms the osmotic origin of the effective diffusivity.

The formation and propagation of sharp interfaces observed in microfluidic experiments ~\cite{ref3,ref4,ref5,ref6,ref7,ref8,ref9,ref10,ref11,ref13,ref14,ref15,ref16,ref17,ref18,ref19}, together with the critical $Pe$ controlled by $D_{eff}$, are in direct agreement with the theoretical framework developed in this study. This quantitative agreement between theory and experiment confirms the osmotic origin of the effective diffusivity. Beyond this scaling view, the framework also identifies a microscopic criterion for interface transparency. We denote by $E_b$  the barrier energy of the FD interface, obtained from the quadratic variation of the free-energy functional, $E_{b} = \langle b|H|b \rangle$ with $\qquad H = -\kappa \partial_{z}^{2} + h''[c_{0}(z)]$. Within the same framework, the critical driving force reduces to the simple relation $\Delta \mu_{\text{crit}} = E_{b}$, directly linked to grand-canonical fluctuations, whose derivation is straightforward and therefore omitted here.

We have presented a theoretical framework that explains how sharp interfaces can persist in fully miscible fluids without invoking bulk phase separation or a double-well potential. By deriving a free energy functional that yields a Fermi–Dirac–type profile as the exact solution under local chemical-potential equilibrium, we identified a class of interfaces that are thermodynamically stable, spatially sharp, and analytically tractable. Under a small imposed chemical-potential drop $\Delta\mu$, these interfaces translate rigidly with the interfacial chemical-potential gradient scaling as $O(\Delta\mu^{2})$, resulting in vanishing leading-order dissipation and a flux entirely determined by the bulk effective diffusivity. Scaling analysis shows that this reduced-resistance property persists over a broad range of $\Delta\mu$ before transitioning to an interface-limited regime. The formation of the FD interface follows an irreversible pathway governed by Onsager’s least dissipation principle, establishing a locally dissipation-free sliding equilibrium. Over longer timescales, irreversible energy dissipation induced by osmotic drift increases the total entropy, driving interfacial ripening and ultimately the breakdown of the FD profile, fully consistent with nonequilibrium thermodynamics. In sharp contrast, the Fickian diffusion interface generates divergent entropy production at contact, representing the most dissipative path rather than the least. This mechanism therefore establishes least-dissipation interfaces as a universal alternative to Fickian diffusion, overturning the conventional view that convexity precludes interfacial structures. It offers a unified, predictive description of sharp interfaces across diverse fields, including polymers, colloids, biological transport, and marine systems.

\begin{acknowledgments}
H.S.L. acknowledge the support from Dong-A University; the Technology Innovation Program (No. 20010315), the Ministry of Trade, Industry \& Energy (MOTIE, Korea). the Ministry of Trade, Industry \& Energy (MOTIE) of the Republic of Korea (RS-2023-00258521). Basic Science Research Program through the National Research Foundation of Korea (NRF) funded by the Ministry of Education (RS-2023-00248126).
\end{acknowledgments}

\nocite{*}

\bibliography{apssamp}

\end{document}